# Benchmarking Open-Source Speech Emotion Recognition in Naturalistic Mandarin Spine Clinic Consultations: A Pilot Validation Study

Tsz Yuet Yeung[1], Zonglin He[1], Dong Chen[1,2], Huili Peng[2], Huiren Tao[1,2], Kenneth MC Cheung[1,2]

[1]The LKS Faculty of Medicine, The University of Hong Kong, Hong Kong SAR

[2]The University of Hong Kong-Shenzhen Hospital, Shenzhen

**Abstract**

Background: Speech emotion recognition (SER) offers a potential means for passive and scalable monitoring of affect during clinical encounters, but most systems are validated on acted laboratory speech rather than naturalistic Mandarin outpatient consultations. Therefore, this study aimed to benchmark open-source SER models against a researcher-consensus reference standard in naturalistic spine-clinic speech and to evaluate whether accuracy reflects clinically meaningful detection of minority affective states under class imbalance.

Methods: In this retrospective analysis, prospectively collected single-center consultation audio recordings were used to compare three open-source SER architectures (emotion2vec+, SenseVoice, FunASR). The audio recordings were first preprocessed to ensure standardized volume levels, and only the conversations between patients, patients' family and clinicians were included. A total of 65 utterances (5–50 s each; one per participant; 31 patients, 34 family members) were analyzed. Six qualified annotators assigned six-category labels (Happy, Sad, Fear, Anger, Neutral, Surprised) after calibration; consensus labels were obtained via majority voting with Fleiss' κ filtering and clinician adjudication for low-agreement segments. We evaluated, reporting unweighted accuracy (UA), macro-average per-class accuracy, class-level and sample-level weighted accuracy (WA), and F1 with bootstrap 95% CIs.

Results: Reference labels were imbalanced (Neutral 58.5%). Median Fleiss' κ was 0.230 (IQR 0.134–0.519). SenseVoice and FunASR achieved an UA of 61.5% (95% CI 49.2–73.8%), a Macro-Average Per-Class Accuracy of 87.2%, and divergent WA (Class-level= 92.5%, Sample-level=24.6%). emotion2vec+ yielded lower UA of 55.4% (95% CI 42.5–67.7%) and Macro-Average Per-Class Accuracy of 85.1%, with 90.6% Weighted Class-Level Accuracy, and 24.2% Weighted Sample Accuracy. All models exhibited near-zero recall for Sad, Fear, Anger, and Surprised, despite demonstrating a high inter-model agreement of 90.8%.Conclusions: In this exploratory pilot, majority-class accuracy was misleading: open-source SER showed poor macro recall for minority emotions against a noisy naturalistic reference. Deployment readiness for affective monitoring cannot be inferred from acted-corpus benchmarks without domain adaptation, multimodal modeling, stronger reference standards, and validation against clinical outcomes.

## Introduction

Effective communication is the cornerstone of the therapeutic alliance, with verbal interactions serving as the primary means of maintaining the clinician-patient relationship and the alliance (1, 2). During clinical consultations, linguistic text is often insufficient for interpreting a patient's affective state. Alternatively, subtle paralinguistic cues, including micro-variations in pitch, speech rate, and vocal cord tension, carry more authentic, objective emotional data of the patient's affective state (3-5). Accurate recognition of patients' affective states has a great clinical and diagnostic significance. Unaddressed, negative emotions are major drivers for formal patient complaints and medico-legal litigation (6, 7). This diagnostic challenge is exacerbated in high-stress medical environments where patients and their family members seldom explicitly express their psychological distress (4). Instead, they commonly exhibit complex, layered emotions via blunted vocal tones that frontline clinicians may overlook due to time constraints, acute cognitive fatigue, or systemic burnout (8, 9).

While traditional observer-rated scales and patient self-report questionnaires help visualize patients' affective states, these psychometric instruments are subject to significant limitations, including vulnerability to recall bias, clinician subjectivity, and severe operational constraints in fast-paced clinical environments (10, 11). To address these systemic gaps, automated Speech Emotion Recognition (SER) systems have been developed as non-invasive, continuous tools that capture quantifiable vocal biomarkers of emotion.

SER has progressed rapidly from handcrafted acoustic features—fundamental frequency ($f_0$), Mel-Frequency Cepstral Coefficients (MFCCs), and spectral energy (12-14)—to deep learning architectures that learn representations directly from waveforms and spectrograms (15, 16). Self-supervised transformers such as emotion2vec+ (17), unified encoder–decoder systems such as SenseVoice (18), and pipeline frameworks such as FunASR (19) now report strong results on curated benchmarks, including CASIA (20), CNSCED (21), and IEMOCAP. Despite this progress, a critical translational question remains unanswered: *do these systems retain their accuracy when deployed in live clinical environments?* Outpatient consultations differ fundamentally from laboratory recordings—speech is spontaneous rather than scripted, emotions are subtle rather than prototypical, ambient hospital noise is ever-present, and class distributions are heavily skewed toward neutral, task-oriented dialogue (19, 22-24).

We address this question through a prospective clinical validation study at the spine outpatient clinics of a tertiary hospital in Southern China. The objectives of this study are multifold: (1) establish a reproducible capture-and-annotation protocol; (2) report reference-standard reliability explicitly; and (3) benchmark emotion2vec+ and SenseVoice using class-balanced metrics (UA, macro F1, per-class recall) alongside simple baselines.

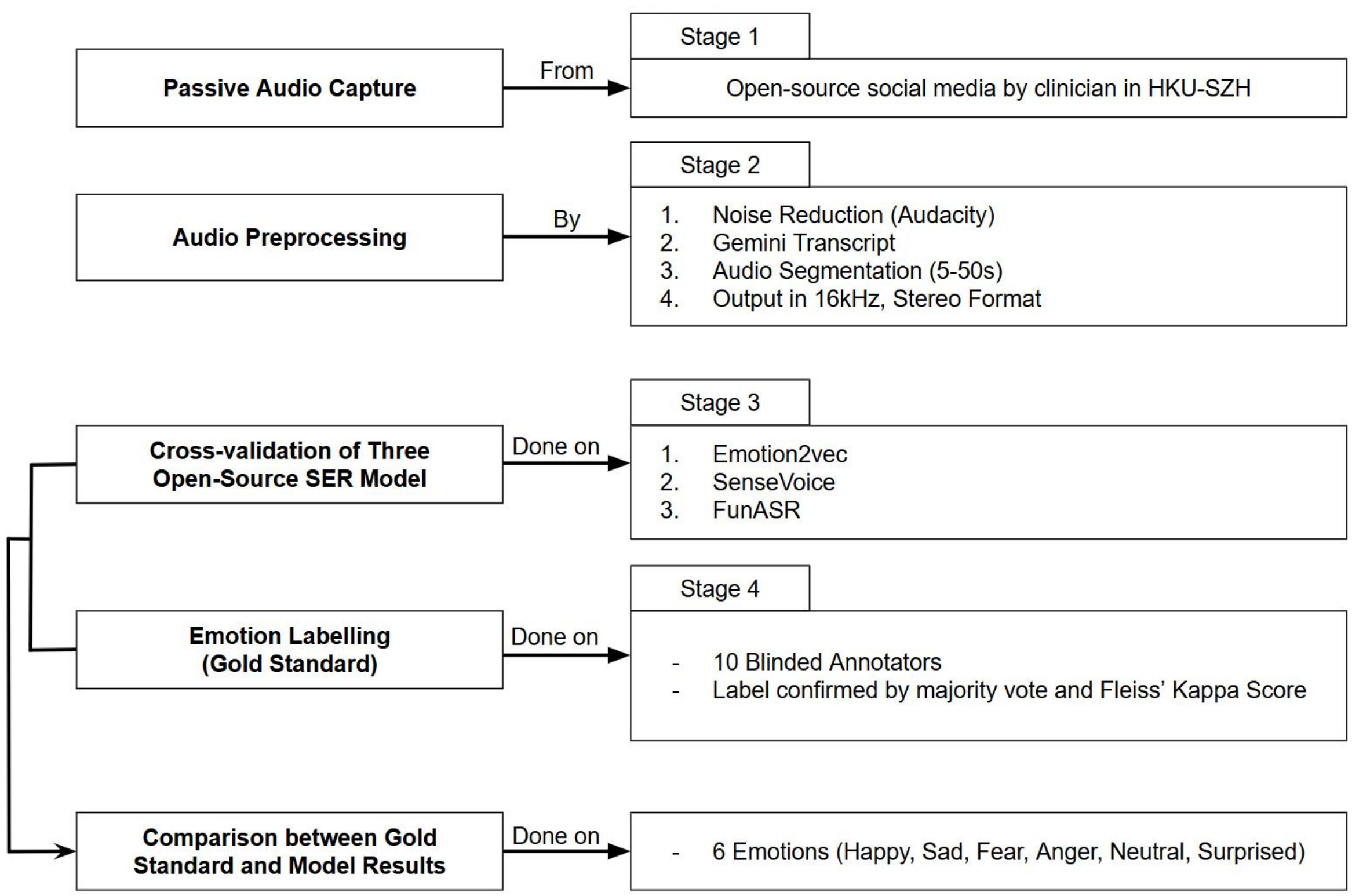

**Figure 1.** Clinical Speech Emotion Recognition validation framework. **Stage 1** passively captures naturalistic spine-clinic consultations (HKU-SZH) via a clinician-worn lapel microphone (32-bit PCM). **Stage 2** applies noise reduction, Gemini transcription, turn segmentation (5–50 s), and 16 kHz stereo export. **Stage 3** cross-validates three open-source SER models: emotion2vec+, SenseVoice, and FunASR. **Stage 4** derives an expert gold standard through 10 blinded annotators, majority vote, and Fleiss' $\kappa$. Predictions are compared against consensus labels across six emotions (Happy, Sad, Fear, Anger, Neutral, Surprise). The bottom panel contrasts actor-recorded laboratory corpora with the naturalistic clinical dataset collected in this study.

**Table 1. Comparison of the currently available Chinese emotion dataset**

| Dataset | Utterances | Total Length | Avg. Length | Dataset Type |
|---|---|---|---|---|
| CASIA | 9,600 | ~8 h | ~2.6 s | Acted: Professional studio recordings |
| CSEMOTIONS | 4,160 | 10.24 h | ~8.0 s | Acted: High-fidelity (48 kHz) professional acting |
| CNSCED | 15,777 | 14 h | — | Natural: TV news and interview programs |
| ESD (Mandarin) | 17,500 | 14.5 h | 3.22 s | Acted: Native Mandarin speakers |
| EmotionTalk | 19,250 | 23.6 h | 4.4 s | Multimodal: Interactive dialogues |
| EMOVIE | 9,724 | 4.18 h | 1.78 s | Multimodal: Extracted from movies |

CNSCED: Chinese Natural Speech Complex Emotion Dataset; ESD: Emotional Speech Dataset.

## Methodology

### *Study Setting and Design*

This prospective, observational clinical validation study was conducted at the University of Hong Kong–Shenzhen Hospital (HKU-SZH) from Jan 2026 to Apr 2026. Figure 1 summarizes the full pipeline. Stage 1 passively captures naturalistic doctor–patient–family consultations via a clinician-worn lapel microphone, preserving the provider's spatial acoustic perspective without disrupting clinical workflow. Stage 2 preprocesses raw naturalistic audio through noise reduction, Gemini-based transcription and timestamping, and turn-level segmentation (5–50 s), with audio exported at 16 kHz. Stage 3 applies three independently evaluated open-source SER models—emotion2vec+ (self-supervised waveform embeddings), SenseVoice (joint ASR/SER encoder–decoder), and FunASR (VAD-coordinated AutoModel pipeline)—to each segment. Stage 4 establishes an expert gold standard through blinded multi-rater annotation, majority-vote consensus, and Fleiss' $\kappa$ filtering, yielding labels across six emotion categories.

### *Participant Recruitment*

Clinical consultations between participants and clinicians were prospectively recorded for a quality improvement study at the Outpatient Clinics of the Spine Center at the University of Hong Kong–Shenzhen Hospital (HKU-SZH) to promote patient education and shared decision-making. Ethical approval has been obtained from the internal review board at HKU–SZH ([2022]252), and patient consent has been sought before recording. Later, the video recordings would be edited to anonymize and de-identify the patients and uploaded to various social media (viz. Xiaohongshu, WeChat Video Platform, TikTok, and Bilibili). Audio from real consultations in these videos was extracted and processed for this study.

Inclusion Criteria are native Mandarin-speaking adults (age >18) admitted to the Spine Center Outpatient Clinics for clinical assessment or follow-up, including patients and their companions. Exclusion Criteria are patients with speech impairments, severe cognitive impairments, and psychological illness.

### *Audio Capture and Preprocessing*

Consultations were recorded with a clinician-worn omnidirectional lapel microphone (DJI; 32-bit PCM MP3), capturing the provider's acoustic perspective. To ensure high-fidelity inputs for downstream SER analysis, raw consultation audio was imported into Audacity Version 3.7.7 (Audacity Team, South Pasadena, the United States) (25), where a background noise profile was captured to apply digital noise reduction, followed by a 10 dB volume amplification to normalize low-amplitude speech across participants. The processed audio was then processed via the Gemini 3.5 Flash multi-modal API to generate an automated, timestamped transcript of the session. To maintain the conversational structure of the interaction while isolating relevant clinical signals, dialogue segments were filtered using timestamps. Patients' or their family members' utterances that last longer than 2 seconds are preserved, with a subsequent clinician response, to maintain conversational context and to strip out short verbal fillers. Annotators labeled audio segments; access to the transcript during lawbeling was limited to calibration and was not used as a model input. Segments containing patient/family speech (>2 s) with subsequent clinician context were exported as dual-channel 16 kHz stereo. This configuration aligns with established benchmarks in the machine learning industry for downstream SER feature extraction (26).

***Reference Standard Mapping***

Ten trained researchers independently reviewed each segment after passing the ESD/CASIA-based qualification. Up to six qualified ratings per utterance contributed to the majority vote. Labels comprised six basic emotions used in SER benchmarks (Happy, Sad, Fear, Anger, Neutral, and Surprised). We acknowledge that clinical spine consultations may express distress, worry, or pain-related flat affect that do not map cleanly onto acted basic-emotion categories; this taxonomy was chosen for comparability with open-source SER outputs. Fleiss' κ was computed per utterance for nominal multi-rater agreement (27). Segments with $\kappa < 0$ underwent clinician adjudication; final retained labels constitute the reference standard (Table 2).

***SER Models***

Three open-source SER platforms were selected for their public availability and native support for Mandarin Chinese speech. Specifically, emotion2vec+ utilizes a self-supervised online distillation pipeline that simultaneously minimizes frame-level and utterance-level losses during pre-training. It ingests raw waveforms, bypasses traditional handcrafted feature engineering, and maps the signals into dense, low-dimensional affective embeddings. These embeddings are subsequently classified across fine-grained discrete emotion categories via a lightweight downstream linear layer. Proving highly resilient across diverse multilingual contexts, emotion2vec+ consistently achieves weighted F1-scores ranging from 48.7% to 84.45% on major benchmarks such as the Ryerson Audio-Visual Database of Emotional Speech and Song (RAVDESS) (28), effectively decoupling emotional prosody from linguistic and speaker-specific variations (17).

SenseVoice implements a unified encoder-decoder architecture that ingests an 80-dimensional log-mel filter-bank. Rather than relying on separate downstream classification heads, it processes the audio token streams to execute Automatic Speech Recognition (ASR), Language Identification (LID), Audio Event Detection (AED), and SER simultaneously. On cross-corpus validation tests using standardized datasets such as Interactive Emotional Dyadic Motion Capture (IEMOCAP) (29), SenseVoice-Small achieves a Weighted Accuracy (WA) of 65.7% and a macro F1-score of 67.9% without target-domain fine-tuning (18).

FunASR operates using a centralized pipeline model configuration (AutoModel). When processing an audio stream, it sequentially coordinates Voice Activity Detection (via FSMN-VAD) to isolate human speech, routes the segments to specialized extraction modules, and concurrently maps downstream paralinguistic elements (30).

**Accuracy metrics**

Given the severe class imbalance, a single accuracy metric may obscure differential performance across emotion classes. We therefore computed four complementary accuracy metrics to benchmark the model performances, as reported elsewhere (31, 32):

(1) Unweighted Accuracy: the proportion of correctly classified samples.
(2) Macro-Averaged Per-Class Accuracy: the unweighted mean of per-class accuracies, giving equal

importance to each emotion of the emotion classes (n=6) regardless of sample size.
(3) Weighted Class-Level Accuracy: $sum(w_i*(TP_i+TN_i)) / (N * sum(w_i))$
(4) Weighted Sample Accuracy: An alternative formulation that weights each individual sample's correctness by its class weight, directly answering the question: "what proportion of the total weight was correctly classified?". This is calculated by $sum(w_j * I(correct_j)) / sum(w_j)$.
(5) Composite Score of Accuracy: accuracy + Fleiss' kappa > 0.2 bonus (0/1/2).

Specifically, un*weighted (overall) accuracy* (UA) denotes overall accuracy, and we used two weighted methods to account for the unbalanced classes of emotions in the dataset, where class-weighted accuracy accounts for class imbalances by allocating different weights to each class, while sample-weighted accuracy is based on Fleiss' Kappa ($\kappa$) scores. A composite score was derived from the original dataset by combining prediction correctness with the Fleiss' $\kappa > 0.2$ criterion. For each model and sample, the score takes one of three values: 0 (neither correct nor $\kappa > 0.2$), 1 (either correct or $\kappa > 0.2$), or 2 (both correct and $\kappa > 0.2$).

**Statistical Analysis**
All analyses were performed in R (version 4.3.1). Statistical significance was assessed at alpha = 0.05. Fleiss' Kappa was computed per sample to quantify the reliability of the gold-standard labels. Kappa values were summarized descriptively and stratified by emotion class. Baselines included always predicting Neutral (prevalence-matched) and uniform random guessing (UA = 16.7%). Given a median $\kappa = 0.230$, model performance is interpreted relative to the reference-standard noise and baselines.
Class weights were computed as the inverse of class frequency, such that minority classes received proportionally higher weights to counteract the dominance of the majority class. For each sample, the inter-annotator agreement among the human raters who produced the gold-standard labels was quantified using Fleiss' Kappa ($\kappa$). Samples with $\kappa > 0.2$ were considered to have acceptable agreement, a threshold commonly used in behavioral research.

To test whether the three models' unweighted accuracy rates differed, we used Cochran's Q test, a non-parametric test for three or more paired binary outcomes. Post-hoc pairwise comparisons were conducted using McNemar's test with continuity correction, appropriate for paired binary data, with a Bonferroni-corrected significance threshold of $\alpha / 3 = 0.0167$ to control the family-wise error rate.

To compare the sample-level weighted scores across the three models, we used the Friedman test, a non-parametric alternative to repeated-measures ANOVA for k related samples. Effect size was quantified using Kendall's coefficient of concordance (W). Pairwise post-hoc comparisons were conducted using the Wilcoxon signed-rank test with Bonferroni correction.

To examine whether model performance varied as a function of annotation quality and speaker context, we conducted subgroup analyses stratified by two binary factors: (a) Fleiss' Kappa (kappa >= 0.2 vs. kappa < 0.2), representing high- versus low-agreement samples, and (b) Speaker type (Patient vs. Family), representing the two speaker groups present in the dataset. For each of the four resulting subgroups, we computed all four accuracy metrics (Unweighted, Macro-Averaged, Weighted Class-Level Accuracy, Weighted Sample-Level Accuracy) for each model. Within each subgroup, Cochran's Q test was applied to the unweighted accuracy data to test model differences. Two-by-three contingency tables (Correct/Incorrect x three models) were constructed for each subgroup to describe the unweighted accuracy patterns.

**Results**
Naturalistic audio recordings of consultations from 65 participants were collected and analyzed in this study. The cohort was approximately equally divided between patients (n = 31) and their family members (n = 34). Within the patient subgroup, the demographic distribution consisted of 17 female and 14 male participants. Conversely, the family member subgroup exhibited a higher proportion of females (n = 23) than males (n = 11). Of the 65 samples, 39 (60.0%) exceeded the κ>0.2 threshold (mean κ = 0.336, SD = 0.360, range [-0.348, 1.000]).

Fleiss' Kappa varied substantially across gold-standard emotion classes. Agreement was highest for Happy (kappa = 0.760, SD = 0.264), followed by Neutral (kappa = 0.351, SD = 0.338), Sad (kappa = 0.315, SD = 0.429), and Anger (kappa = 0.230, SD = 0.000). Agreement was poor for Fear (kappa = 0.076, SD = 0.242) and Surprised (kappa = 0.013, SD = 0.213). This suggests that the gold-standard labels themselves are least reliable for the minority classes that the models also fail to recognize.

**Table 1. Study cohort, annotation quality, and emotion label characteristics.**

| Domain | Metric | Total | Happy | Sad | Neutral | Anger | Fear | Surprised |
|---|---|---|---|---|---|---|---|---|
| **Cohort** | Utterances, N (%) | 65 (100%) | 6 (9.2%) | 9 (13.8%) | 38 (58.5%) | 3 (4.6%) | 5 (7.7%) | 4 (6.2%) |
| | Female / Male | 40 / 25 | 3 / 3 | 7 / 2 | 22 / 16 | 2 / 1 | 4 / 1 | 2 / 2 |
| | Patient / Family | 31 / 34 | 2 / 4 | 7 / 2 | 16 / 22 | 3 / 0 | 1 / 4 | 2 / 2 |
| **Audio** | Mean duration (s) | 17.7 | 18.1 | 23.3 | 15.7 | 17.7 | 17.8 | 23.0 |
| | Total duration (s) | 1,149 | 109 | 210 | 596 | 53 | 89 | 92 |
| | Avg. words per audio | 78.1 | 75.9 | 108.4 | 69.8 | 88.0 | 77.0 | 86.5 |
| **Annotation** | Qualified annotators | 6 | / | / | / | / | / | / |
| | Median κ (IQR) | 0.230 (0.134 to 0.519) | 0.759 (0.519 to 1.000) | 0.134 (-0.059 to 0.519) | 0.230 (0.134 to 0.519) | 0.230 (0.230 to 0.230) | 0.134 (0.134 to 0.230) | 0.037 (-0.107 to 0.158) |
| | κ < 0.2, N (%) | 26 (40.0%) | 0 (0.0%) | 5 (55.6%) | 15 (39.5%) | 0 (0.0%) | 3 (60.0%) | 3 (75.0%) |
| | κ < 0, N (%) | 14 (21.5%) | 0 (0.0%) | 3 (33.3%) | 8 (21.1%) | 0 (0.0%) | 1 (20.0%) | 2 (50.0%) |
| **Weight** | Class weight | / | 1.806 | 1.204 | 0.285 | 3.611 | 2.167 | 2.708 |

*Unweighted and Macro-Averaged Accuracy*

Unweighted accuracy was numerically highest for SenseVoice and FunASR (both 40/65 = 61.5%, 95% CI [48.6%, 73.3%]), followed by Emotion2Vec+ (36/65 = 55.4%, 95% CI [42.5%, 67.7%]; Table 2). Cochran's Q test did not reach statistical significance (Q = 5.33, df = 2, p = 0.070), suggesting no reliable difference in unweighted accuracy among the three models. Pairwise McNemar tests were likewise non-significant after Bonferroni correction (Emotion2Vec+ vs. SenseVoice: $\chi 2$ = 1.50, p = .221; Emotion2Vec+ vs. FunASR: $\chi 2$= 1.50, p = .221). SenseVoice and FunASR produced identical predictions on all 65 samples, precluding a meaningful pairwise comparison.

Macro-averaged per-class accuracy, which gives equal weight to all six emotion classes, was substantially higher than unweighted accuracy across all models: 85.1% (Emotion2Vec+), 87.2% (SenseVoice), and 87.2% (FunASR).

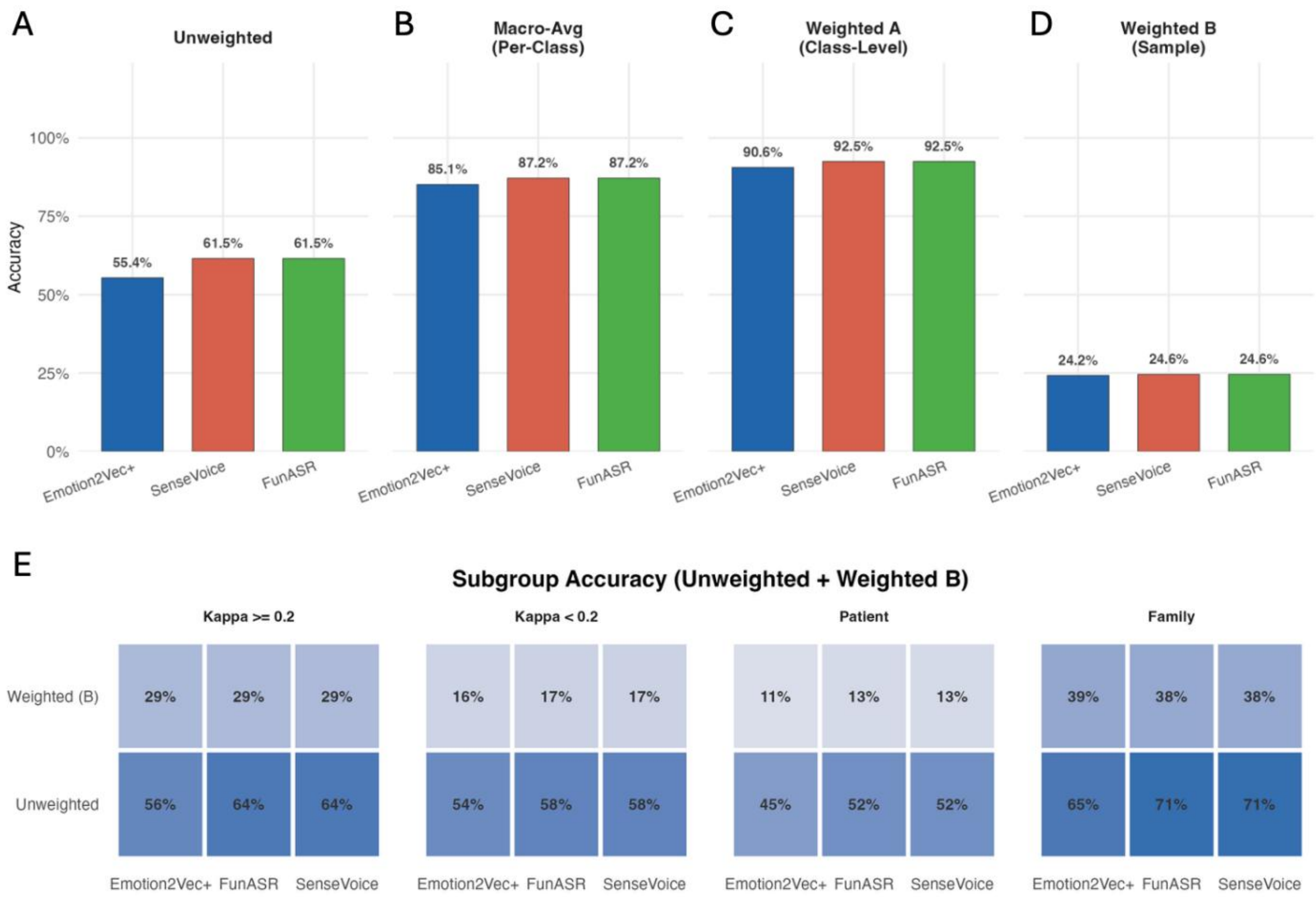

Figure 2. Model accuracy across four metrics. (A) unweighted; (B) macro-average per-class; (C) class-weighted; (D) sample-weighted accuracy metrics; (E) Subgroup accuracy.

*Weighted Accuracy*

The two weighted accuracy measures produced markedly divergent results, as shown in Table 2. Weighted Class-Level Accuracy (Method A) yielded high accuracies of 90.6% (Emotion2Vec+) and 92.5% (SenseVoice and FunASR. In contrast, Weighted Sample-Level Accuracy (Method B) was substantially lower than all other metrics, with 24.2% for Emotion2Vec+ and 24.6% for SenseVoice and FunASR.

**Table 2. Accuracy performance of three models.**

| Model | Correct | N | Unweighted | 95% CI | Macro-Avg Per-Class | Weighted (A) Class-Level | Weighted (B) Sample | Composite (0-2) |
|---|---|---|---|---|---|---|---|---|
| Emotion2Vec+ | 36 | 65 | 0.554 | [0.425, 0.677] | 0.851 | 0.906 | 0.242 | 1.154 |
| SenseVoice | 40 | 65 | 0.615 | [0.486, 0.733] | 0.872 | 0.925 | 0.246 | 1.215 |

| Model | Correct | N | Unweighted | 95% CI | Macro-Avg Per-Class | Weighted (A) Class-Level | Weighted (B) Sample | Composite (0-2) |
|---|---|---|---|---|---|---|---|---|
| FunASR | 40 | 65 | 0.615 | [0.486, 0.733] | 0.872 | 0.925 | 0.246 | 1.215 |

*Note.* Composite score range: 0 to 2 (prediction correctness + Fleiss' κ > 0.2 bonus).

The Friedman test on sample-weighted scores was not significant ($\chi2 = 5.33$, df = 2, p = .070; Kendall's W = 0.041), and pair-wise Wilcoxon signed-rank tests were non-significant after Bonferroni correction (Emotion2Vec+ vs. SenseVoice: V = 6, p = 0.374; Emotion2Vec+ vs. FunASR: V = 6, p = 0.374).

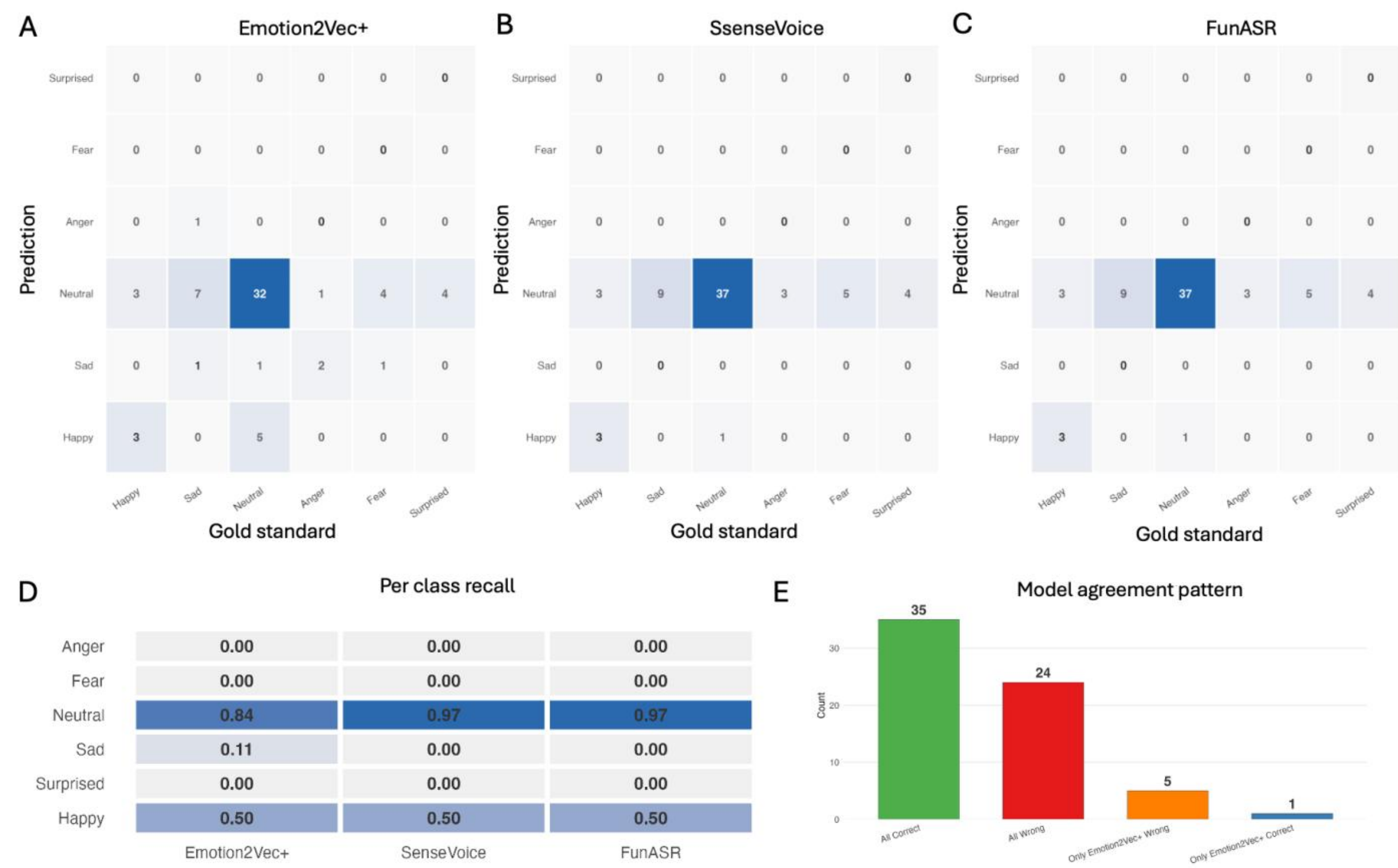

Figure 3. Confusion metrics and recall rates of three models.

*Per-Class Performance*

All three models exhibited a pronounced performance disparity across emotion classes (Figure 2). The majority class, Neutral (n = 38, weight = 0.285), was recognized with high recall by all models (SenseVoice and FunASR: 97.4%; Emotion2Vec+: 84.2%). Happy (n = 6, weight = 1.806) was detected at 50.0% recall by all three models. Performance on the remaining four classes was poor or absent: Sad was detected at 11.1% recall by Emotion2Vec+ and 0% by the other two models; Anger, Fear, and Surprised were not detected by any model (0% recall). Importantly, these four undetected or poorly detected classes carry the highest weights (Anger: 3.611; Surprised: 2.708; Fear: 2.167; Sad: 1.204), directly explaining the low weighted sample accuracy (Method B) of approximately 24%.

*Patient vs Family Speakers*

Among patient speakers (n = 31), SenseVoice WA was 61.3% and UA 16.7%; non-Neutral recall was 0.0% (0/12). Among family speakers (n = 34), WA was 61.8% and UA 28.3%; non-Neutral recall was 20.0% (3/15). Given small cell sizes, these differences are hypothesis-generating only.

Table 3. Comparative analysis and subgroup analysis

| Section / Metric | Emotion2 Vec+ | SenseVoic e | FunASR | Statistic | df | p-value |
|---|---|---|---|---|---|---|
| **A. Overall Comparisons (N = 65)** | | | | | | |
| | **Emotion2 Vec+** | **SenseVoic e** | **FunASR** | **Statistic** | **df** | **p-value** |
| Unweighted Acc | 55.4% | 61.5% | 61.5% | | | |
| Macro-Avg Per-Class | 85.1% | 87.2% | 87.2% | | | |
| Weighted (A) Class-Level | 90.6% | 92.5% | 92.5% | | | |
| Weighted (B) Sample-Level | 24.2% | 24.6% | 24.6% | | | |
| Composite Score (0-2) | 1.15 | 1.22 | 1.22 | | | |
| | | | | | | |
| Cochran's Q | | | | Q=5.333 | 2 | 0.0695 |
| Friedman test | | | | $\chi2$=5.333 | 2 | 0.0695 |
| Kendall W = 0.041 | | | | | | |
| | | | | | | |
| **Post-hoc (Bonferroni alpha=0.0167)** | | | | **McNemar $\chi2$** | | **Wilcoxon V** |

| Section / Metric | Emotion2 Vec+ | SenseVoic e | FunASR | Statistic | df | p-value |
|---|---|---|---|---|---|---|
| EV2+ vs SenseVoice | | | | 1.500 | 0.2207 | V=6, p=0.3741 |
| EV2+ vs FunASR | | | | 1.500 | 0.2207 | V=6, p=0.3741 |
| SenseVoice vs FunASR | | | | NaN | NaN | V=0, p=NaN |
| **B. Subgroup Analyses** | | | | | | |
| Subgroup (N) | Metric | EV2+ | SV | FS | Cochran Q | Cochran p |
| Kappa >= 0.2 (n=39) | Unweighte d Acc | 56.4% | 64.1% | 64.1% | Q=3.60 | 0.1653 |
| | **Correct (n)** | **22** | **25** | **25** | | |
| | **Incorrect (n)** | **17** | **14** | **14** | | |
| | **Weighted (B) Acc** | **29.3%** | **29.2%** | **29.2%** | | |
| Kappa < 0.2 (n=26) | Unweighte d Acc | 53.8% | 57.7% | 57.7% | Q=2.00 | 0.3679 |
| | **Correct (n)** | **14** | **15** | **15** | | |
| | **Incorrect (n)** | **12** | **11** | **11** | | |
| | **Weighted (B) Acc** | **16.0%** | **17.2%** | **17.2%** | | |
| Patient (n=31) | Unweighte d Acc | 45.2% | 51.6% | 51.6% | Q=4.00 | 0.1353 |
| | **Correct (n)** | **14** | **16** | **16** | | |
| | **Incorrect (n)** | **17** | **15** | **15** | | |

| Section / Metric | Emotion2 Vec+ | SenseVoic e | FunASR | Statistic | df | p-value |
|---|---|---|---|---|---|---|
| | **Weighted (B) Acc** | **11.4%** | **13.0%** | **13.0%** | | |
| Family (n=34) | Unweighte d Acc | 64.7% | 70.6% | 70.6% | Q=2.00 | 0.3679 |
| | **Correct (n)** | **22** | **24** | **24** | | |
| | **Incorrect (n)** | **12** | **10** | **10** | | |
| | **Weighted (B) Acc** | **39.2%** | **38.0%** | **38.0%** | | |

*Inter-Model Agreement and Composite Scores*

The three models exhibited very high agreement: on 35 of 65 samples (53.8%), all three models were simultaneously correct, and on 24 samples (36.9%), all three were simultaneously incorrect, yielding 90.8% unanimous agreement. In only 5 samples Emotion2Vec+ was the sole model to err (while SenseVoice and FunASR were correct), and in a single sample, Emotion2Vec+ was the only correct model. SenseVoice and FunASR were perfectly concordant across all 65 samples.

Composite scores (accuracy + Fleiss' κ > 0.2 bonus; range 0-2) followed the same ordinal pattern: Emotion2Vec+ (1.154) scored slightly lower than SenseVoice and FunASR (both 1.215), reflecting the consistent five-sample advantage of the latter two models.

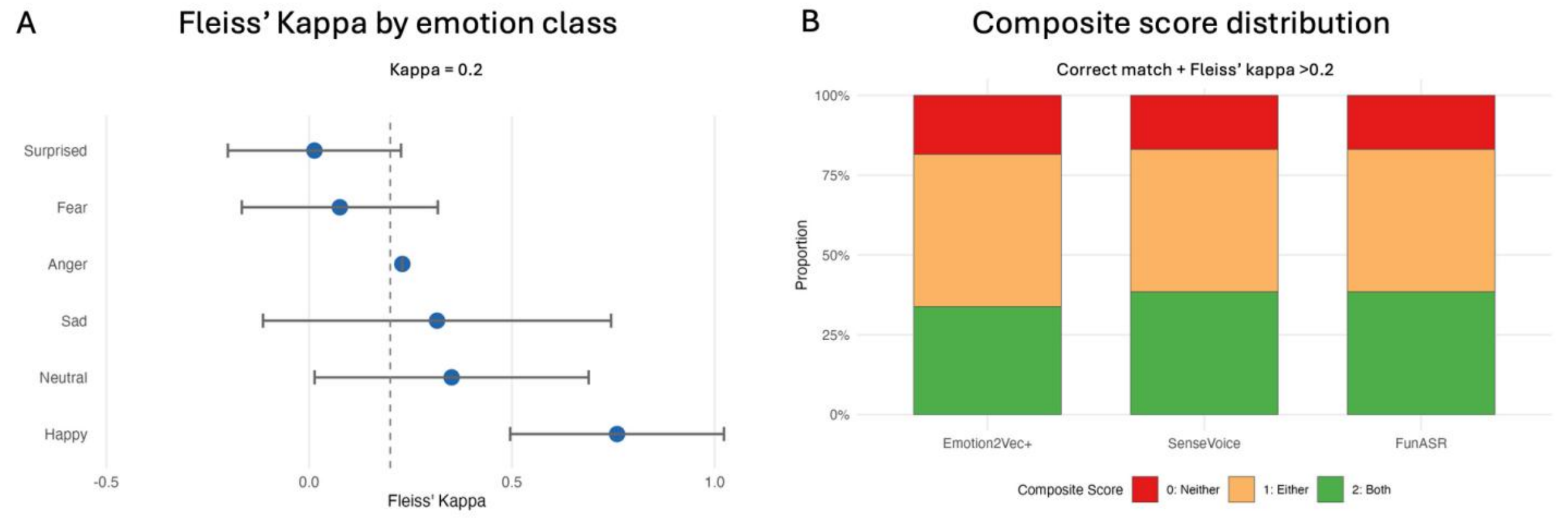

Figure 4. Inter-Model Agreement and Composite Scores

*Subgroup Analysis*

Subgroup analyses stratified by Fleiss' Kappa and speaker type revealed meaningful performance heterogeneity (Table 3, Figure 2). By Kappa stratum, unweighted accuracy was slightly higher in the high-agreement (kappa >= 0.2, n = 39) subgroup (Emotion2Vec+: 56.4%; SenseVoice/FunASR: 64.1%) than in the low-agreement (kappa < 0.2, n = 26) subgroup (53.8% and 57.7%, respectively). More strikingly, weighted sample accuracy showed a larger gap: 29.3% (Emotion2Vec+) and 29.2% (SenseVoice/FunASR) in the high-agreement subgroup versus 16.0% and 17.2% in the low-agreement subgroup, indicating that model accuracy heavily relies on low-weight classes when annotator agreement is poor. Cochran's Q was not significant within either Kappa stratum (kappa >= 0.2: Q = 3.60, p = 0.165; kappa < 0.2: Q = 2.00, p = 0.368).

By speaker type, a larger performance gap emerged. The Family subgroup (n = 34) showed higher unweighted accuracy (64.7% for Emotion2Vec+, 70.6% for SenseVoice/FunASR) than the Patient subgroup (n = 31; 45.2% and 51.6%, respectively). This difference was especially pronounced in weighted sample accuracy: 39.2% (Emotion2Vec+) and 38.0% (SenseVoice/FunASR) for Family versus 11.4% and 13.0% for Patient, suggesting that the Patient speech samples contain a greater proportion of high-weight minority emotions that the models fail to recognize. Cochran's Q was not significant within either speaker subgroup (Patient: Q = 4.00, p = .135; Family: Q = 2.00, p = .368).

## Discussion

### *Primary findings*

This study compared three pre-trained speech emotion recognition models (Emotion2Vec+, SenseVoice, FunASR) on a dataset of 65 samples with six emotion classes. Four complementary accuracy metrics and appropriate nonparametric statistical tests were used to characterize model performance under severe class imbalance.

Four key findings emerged. First, no statistically significant difference was found among the three models by either Cochran's Q (p = 0.070) or the Friedman test (p = 0.070), although both approached the conventional significance threshold. SenseVoice and FunASR were functionally identical on this dataset, producing the same predictions on all 65 samples. Second, all three models exhibited a sharp performance gradient across emotion classes: recognizing the majority class (Neutral) with high accuracy (>84%) while largely or entirely failing to detect four of the six emotion classes (Anger, Fear, Surprised, and -- for two of three models -- Sad). Only Happy was detected at a moderate but suboptimal level (50% recall) by all three models. Third, the choice of an accuracy metric fundamentally alters the narrative about model performance. Unweighted accuracy (55-62%) suggests moderate performance. Macro-averaged per-class accuracy (85-87%) and the class-weighted accuracy (91-93%) suggest near-ceiling performance. In contrast, the sample-weighted metric (24%) indicates that the models correctly classify only about one-quarter of the data when minority classes are appropriately emphasized. The discrepancy arises because unweighted and macro-averaged metrics are dominated by the large Neutral class (via raw counts or via true negatives, respectively). Fourth, subgroup analyses revealed substantial performance heterogeneity, where weighted sample accuracy ranged from ~11-13% in the Patient subgroup to ~38-39% in the Family subgroup, and from ~16-17% in low-agreement samples to ~29% in high-agreement samples. This suggests that Patient speech contains a higher concentration of difficult minority emotion categories and that low annotator agreement may reflect genuinely ambiguous signals that models also fail to resolve.

### *Clinical importance of emotion recognition*

Missed Sad/Fear/Anger detections may delay psychosocial support in chronic pain and surgical follow-up contexts. Still, this study did not link SER outputs to referrals, patient-reported outcomes, or clinician actions. We therefore frame findings as a feasibility and benchmarking warning rather than evidence for or against clinical deployment.

In standardized emotional benchmarks, speech samples are typically recorded under pristine laboratory conditions by professional actors instructed to simulate exaggerated, distinct emotional prototypes (23). Due to the hyper-articulated acoustic boundaries in standardized emotional benchmarks, SER models may misinterpret flattened, low-arousal clinical signals as neutral speech. The models could hardly differentiate between a genuinely calm baseline and a pathologically blunted vocal tone concealing severe emotional distress.

Furthermore, real-world clinical data exhibits severe class imbalance and is heavily contaminated by ambient acoustic artifacts, overlapping speech, and speaker-dependent variations in vocal affect (24). When advanced SER networks encounter a lack of prototypical acoustic cues and severe data sparsity for minor emotional classes, their deep convolutional and transformer-based layers struggle to learn distinct spatiotemporal representations (33-36). Consequently, architecture experiences a pronounced drop in sensitivity across less frequent affective dimensions, failing to replicate the pristine performance baselines established under controlled experimental settings.

The per-class accuracy values entering the Method A formula illustrate precisely why this metric diverges from sample-weighted accuracy. Although the models detected zero Anger, Fear, and Surprised samples (TP = 0), the true negative counts (TN) for these classes ranged from 60 to 62 (out of 65), yielding per-class accuracies of 92-95%. When the high class weights are applied and averaged, these inflated values drive the class-weighted accuracy above 90%.

*Real-world implementation of SER is still difficult*

The cross-validated architectures achieve identical peak UA of approximately 61.5% for SenseVoice and FunASR, alongside a baseline of 55.4% for emotion2vec+; however, their WA sharply decline to roughly 24.6% and 24.2%, respectively. This performance degradation on non-neutral emotional categories—visible in the model vs. gold-standard comparison of Figure 1—highlights a profound vulnerability to real-world deployment, yielding metrics far below those reported on open-source corpora.

An analysis of the distribution of affective states within the collected clinical dataset reveals a pronounced imbalance, with most utterances classified as “Neutral”. This heavy skew toward emotionally neutral expressions is highly consistent with the ecological realities of a standard healthcare environment. In routine medical consultations, such as in an outpatient spine clinic, patients and clinicians primarily engage in objective, transactional dialogue centered on symptom description, diagnostic review, and treatment planning. Empirical evidence confirms that naturalistic medical interviews are inherently dominated by neutral and task-oriented information exchange, while explicit or high-arousal emotional expressions remain relatively infrequent (19, 37, 38).

Furthermore, individuals experiencing chronic or progressive physical symptoms frequently adopt a flattened or suppressed emotional baseline during formal interactions with medical authorities, a behavioral adaptation aimed at maintaining cognitive focus and clearly communicating physical distress to ensure objective information exchange (22). This baseline emotional suppression, combined with the structured nature of clinical questioning, explains why the emotion labels were distributed heavily toward neutrality, leaving secondary affective states like sadness, happiness, and fear as minor, context-specific occurrences.

*The Psycholinguistic Challenge of Mandarin*

Validating SER in Mandarin introduces distinct challenges beyond those encountered in Indo-European languages. As a lexical-tonal language, Mandarin uses four tone contours to distinguish word and emotional meanings (39-43); SER models trained predominantly on non-tonal speech often conflate these mandatory pitch trajectories with emotional prosody (44). In the HKU-SZH setting, consultations further involve English–Mandarin code-switching and regional accent diversity, adding variability that actor-recorded benchmarks do not represent (45). Clinically, Chinese outpatient encounters also differ structurally from Western models: semi-open consultation rooms, high patient throughput, and multi-party dialogues that interweave diagnostic questioning with informal conversation (46-49). An SER system must therefore handle rapid speaker turns, overlapping speech, and low-intensity affect embedded within predominantly neutral clinical discourse—conditions absent from controlled laboratory datasets.

*The Clinical Gap*

Despite rapid advances in algorithms, no naturalistic Mandarin speech dataset has been collected in an active clinical setting. Open-source Chinese emotional speech repositories—CASIA (20), CSEMOTIONS (50), CN-SCED (21)—predominantly feature healthy actors reciting scripted utterances in acoustic chambers (Figure 2). While these corpora offer clean signal-to-noise ratios and enable reproducible benchmarking, they lack the ambient noise, spontaneous turn-taking, and longitudinal context of real consultations. Static psychiatric datasets such as DAIC-WOZ (51, 52) capture only cross-sectional snapshots and remain English-centric. This empirical deficit prevents meaningful assessment of whether benchmark-leading SER architectures can generalize to the ecological conditions outlined in Figure 1.

### ***Strengths and Limitations***

This study makes three contributions. First, we establish a protocol for passively collecting and preprocessing naturalistic Mandarin clinical audio without disrupting outpatient workflow. Second, we deploy a rigorous, blinded annotation pipeline—calibrated against ESD and CASIA exemplars—to produce an expert-consensus gold standard with inter-rater reliability filtering. Third, we provide the first cross-model validation of emotion2vec+, SenseVoice, and FunASR on real clinical speech, quantifying the performance gap between laboratory benchmarks and live deployment. We assume that a majority-vote consensus label from trained, blinded clinical researchers constitutes a valid proxy for ground-truth affect, against which automated predictions are benchmarked for sensitivity, specificity, and clinical utility.

This study has limitations. First, the small sample size and the single-center design limit the generalizability of the findings. Second, the structural efficacy of automated SER within naturalistic outpatient settings is constrained by linguistic, environmental, and unimodal limitations. Owing to the geographic locations, the majority of the subjects were from the Greater Bay Area in China, resulting in an imbalanced distribution of phonological variations. Moreover, phonological variations across Northern and Southern Mandarin, together with English–Mandarin code-switching and regional accent diversity typical of Hong Kong–Shenzhen outpatient settings, alter prosodic baselines and confound acoustic classifiers. Simultaneously, transient clinical noise profiles—such as overlapping medical chatter and equipment alerts—risk masking subtle vocal cues or causing them to be accidentally erased during preprocessing. Because unimodal feature extraction provides a less comprehensive representation of affect, this study cross-examines classification variations under independent audio-only channels, highlighting why isolated streams fail to capture the layered interplay of clinical emotional expression. Therefore, future work should prioritize multimodal architectures that jointly model acoustic prosody and transcript semantics, domain-adaptive fine-tuning on localized clinical data, and expanded annotation across longitudinal consultations. Addressing class imbalance through targeted augmentation and cost-sensitive learning will be essential before passive SER can serve as a reliable component of digital phenotyping and behavioral healthcare monitoring in live outpatient settings.

## Conclusion

This pilot study demonstrates that the laboratory-to-clinic gap in SER is not merely incremental but structural. State-of-the-art open-source architectures—emotion2vec+, SenseVoice, and FunASR—achieve moderate unweighted accuracy (55-62%) on naturalistic Mandarin clinical speech yet collapse on weighted metrics (24%) when rare emotion classes and ecological noise are accounted for. By establishing the first naturalistic Mandarin clinical dataset and a reproducible four-stage validation framework, we provide both an empirical baseline and a roadmap for clinically grounded SER research.

## Supplementary materials

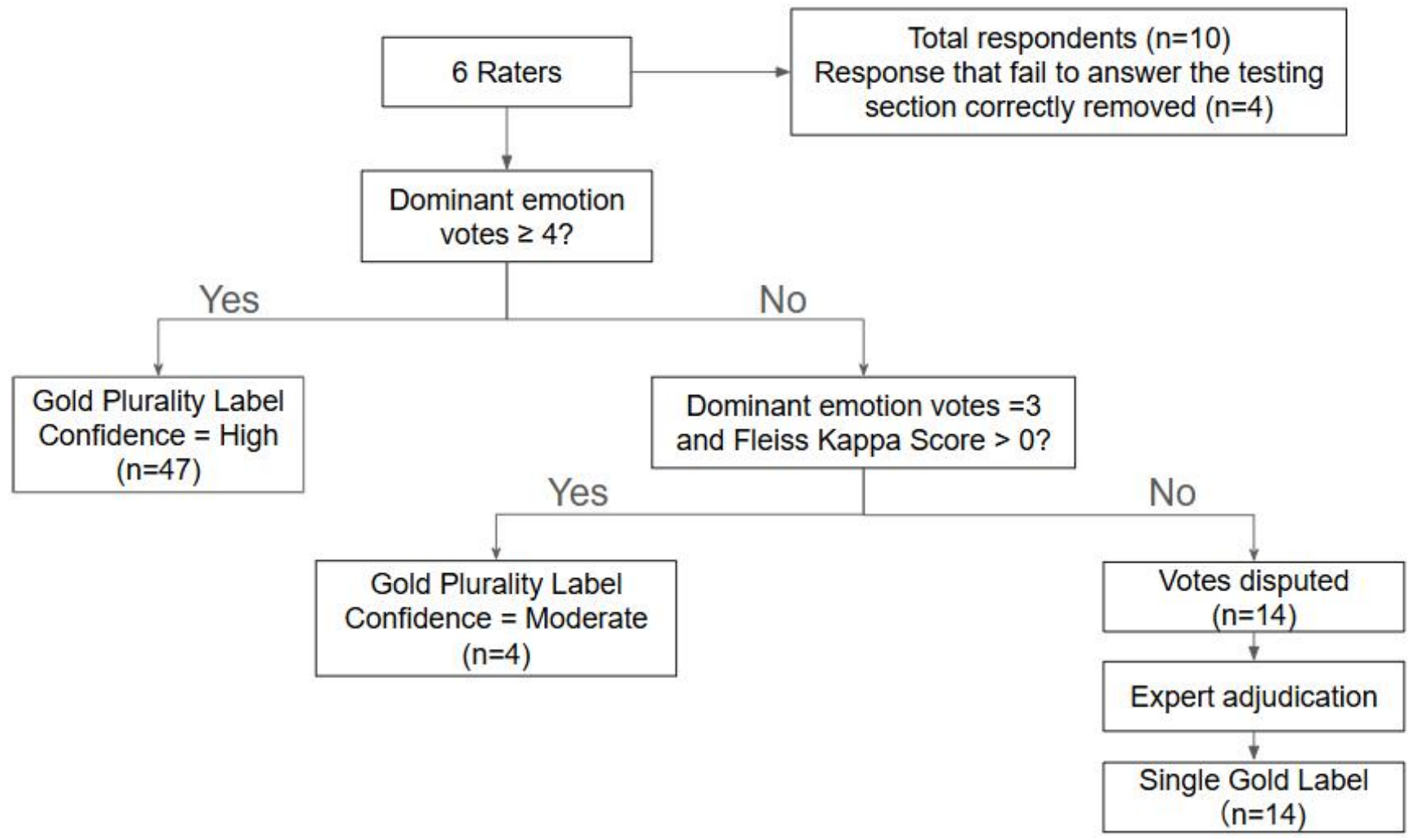

**Figure S1.** Emotion Labeling Process of Audio